\documentclass[twocolumn,showpacs,preprintnumbers,amsmath,aps,amssymb]{revtex4}
\usepackage{graphicx}% Include figure files
\usepackage{dcolumn}% Align table columns on decimal point
\usepackage{bm}% bold math

\begin{document}

\title{WKB-type Approximation to Noncommutative Quantum Cosmology}
\author{E. Mena}
 \email{emena@fisica.ugto.mx}
 \author{O. Obregon}
\email{octavio@fisica.ugto.mx}
\author{M. Sabido}
\email{msabido@fisica.ugto.mx}
\affiliation{ Instituto de F\'{\i}sica de la Universidad de Guanajuato,\\
 A.P. E-143, C.P. 37150, Le\'on, Guanajuato, M\'exico\\
 }%

\date{\today}% It is always \today, today,
\begin{abstract}
In this work, we develop and apply the WKB approximation to  several examples of noncommutative quantum cosmology, obtaining the time evolution of the noncommutative universe, this is done starting from a noncommutative quantum formulation of cosmology where the noncommutativity is introduced by a deformation on the minisuperspace variables. This procedure gives a straightforward algorithm to incorporate noncommutativity to cosmology and inflation.
 \end{abstract}
 \maketitle
\section{Introduction}
There has been a lot of interest in the old idea of noncommutative space-time \cite{snyder}, and an immense amount of work has been done on the subject. This renewed interest is a consequence  of the developments in M-Theory and String Theory, where in the description of the low energy excitations of
open strings, in the presence of a Neveu-Schwarz (NS) constant background $B-$field, a noncommutative effective low
energy gauge theory  action \cite{connes,ref3} appears in a natural way. 
Along the lines of noncommutative gauge theory, noncommutative theories of gravity have been constructed \cite{ncsdg}. All of these formulations of gravity on noncommutative space-time are highly nonlinear and a direct calculation of cosmological models is incredibly difficult. One may naively expect  noncommutative effects to be present at the Planck scale so it is almost impossible to detect them, but due to the UV/IR mixing \cite{min,doug}, the effects of noncommutative might be important at the cosmological scale.

A simplified approach to study the very early universe is quantum cosmology (QC), which means that the gravitational and matter variables have been reduced to a finite number of degrees of freedom (these models were extensively studied by means of Hamiltonian methods in the 
1970's, for reviews see \cite{ryan,maccallum}); for homogenous cosmological models the metric depends only on time, this permits to integrate the space dependence and obtain a model with a finite dimensional configuration space, {\it minisuperspace}, whose variables are the 3-metric components. Unfortunately this construction is plagued with several drawbacks, there is no easy way to study the dynamical evolution of the system; also the wave functions usually can not be normalized.  One way to extract useful dynamical information is through a WKB type method.

In the last few years there have been several attempts to study the possible effects of 
noncommutativity in the classical cosmological scenario \cite{romero,brand}.
 In \cite{ncqc} the authors avoid the difficulties of analyzing noncommutative cosmological models,
  that would arise when working with a noncommutative theory of gravity \cite{ncsdg}. 
  Their proposal introduces the effects of noncommutativity at the quantum level,
   namely quantum cosmology, by deforming the minisuperspace through a Moyal deformation of the Wheeler-DeWitt equation.
     It is then possible to proceed as in noncommutative quantum mechanics \cite{gamboa}.
     On the other hand, noncommutativity among the fields is a consequence of the usual noncommutative space-time \cite{ref3,Jurco:2001rq}
     also for gravitational fields \cite{ncsdg}. So the mentioned proposal is an effective noncommutativity in quantum cosmology.

The aim of this paper is to apply a WKB type method to noncommutative quantum cosmology, 
and find the noncommutative classical solutions, avoiding in this way the difficult task to solve these cosmological models in the complicated framework of noncommutative gravity \cite{ncsdg}. 
 We know how to introduce noncommutativity at a quantum level, by taking into account the changes 
 that the Moyal product of functions induces on the quantum equation (i.e. Schr\"{o}edinger equation),
  and from there calculate the effects of noncommutativity at the classical level.
   This  also has the advantage that for some noncommutaive models for which the quantum solutions 
   can not be found, the corresponding noncommutative classical solutions arise very easily from this formulation. 
   The procedure is presented through a series of examples:
    first the Kanstowski-Sachs cosmological (KS) model is presented in detail, and the formalism developed for this model,
     is then applied to the Friedmann-Robertson-Walker (FRW) universe coupled to a scalar field $\phi$ and cosmological constant $\Lambda$,
      the noncommutative quantum and classical solutions for different cases are presented. The method has the 
      advantage that broadens the noncommutative cosmological models that can be solved. 
      This is of particular interest in conection with inflation; in \cite{brand} the effects of noncommutativity during 
      inflation are explored, but noncommutativity is only incorporated to the scalar field neglecting 
      the gravitational sector, using the method developed in this paper, noncommutativity 
     can be incorporated on both sectors. Finally the procedure is applied to a toy model of string effective quantum cosmology \cite{nuevo}.

This work is organized as follows. In section 2 we review several quantum cosmological models via the Wheeler-DeWitt equation (WDW) and find the corresponding wave function for most of these models, then we obtain the classical solutions using a WKB type approximation. In section 3 we repeat the same treatment to the noncommutative counterparts of the examples presented in section 2, this is achieved through a noncommutative deformation of the minisuperspace variables. Finally, the last section is devoted to discussion and outlook.
%%%%%%%%%%%%%%%%%%%%%%%%
%%%%%%%% Seccion 2
%%%%%%%%%%%%%%%%%%%%%%%%
\section{Quantum Cosmology and the WKB approximation}
Our goal is to present a WKB type method for noncommutative quantum cosmology. We start by reviewing the quantum cosmological models in which we are interested, and find the classical evolution through a WKB type approximation. The models presented are: Kantowski-Sachs cosmology, FRW cosmology with cosmological constant  coupled to a scalar field, and a cosmological model in the framework of string theory.
%%%%%%%%%%%%%%%%%
\subsection{Kantowski-Sachs Cosmology}
The first example we are interested is the Kantowski-Sachs universe, this is one of the simplest anisotropic cosmological models. 
The Kantowski-Sachs line element is \cite{ncqc} 
\begin{equation}
ds^{2}=-N^{2}dt^{2}+e^{ 2\sqrt{3}\beta}  dr^{2}+ e^{-2\sqrt{3}(\beta+\Omega)} \left(d\vartheta^{2}+\sin^{2}\vartheta d\varphi^{2}\right),
 \label{KS metric}%
\end{equation}
from the general relativity Lagrangian  we find the canonical momenta,
\begin{equation}
\Pi_{\Omega}=-\frac{12}{N}e^{-\sqrt{3}\beta-2\sqrt{3}\Omega}\dot{\Omega }, ~
\Pi_{\beta}=\frac{12}{N}e^{-\sqrt{3}\beta-2\sqrt{3}\Omega}\dot{\beta },
\label{mks}
\end{equation}
using canonical quantization, and a particular factor ordering, we get the WDW equation, through the usual identifications $\Pi_\Omega=-i\frac{\partial}{\partial\Omega}$ and $\Pi_\beta=-i\frac{\partial}{\partial\beta}$ we get
\begin{equation}
\left[  \frac{\partial^{2}}{\partial\Omega^{2}}-\frac{\partial^{2}}%
{\partial\beta^{2}}-48e^{  -2\sqrt{3}\Omega}  \right]
\psi(\Omega,\beta)=0.
\label{kswdw}
\end{equation}
The solution to this equation is given by
\begin{equation}
\psi=e^{  \pm i\nu\sqrt{3}\beta}  K_{i\nu}\left(  4e^{
-\sqrt{3}\Omega}  \right)  , \label{KSsolution}%
\end{equation}
where $\nu$ is the separation constant and $K_{iv}$ are the modified Bessel functions.

We now proceed to apply the WKB type method. For this we propose the wave function
\begin{equation}
\Psi(\beta,\Omega)\approx e^{i(S_1(\beta)+S_2(\Omega))},
\label{wkb1}
\end{equation}
the WKB approximation is reached in the limit
\begin{equation}
\left|\frac{\partial^2 S_1(\beta)}{\partial\beta^2}\right|<<\left (\frac{\partial S_1(\beta)}{\partial\beta}\right)^2,~
\left|\frac{\partial^2 S_2(\Omega)}{\partial\Omega^2}\right|<<\left (\frac{\partial S_2(\Omega)}{\partial\Omega}\right)^2,
\label{wkb2}
\end{equation}
and gives the Einstein-Hamilton-Jacobi (EHJ) equation
\begin{equation}
-\left (\frac{\partial S_2(\Omega)}{\partial\Omega}\right)^2+\left (\frac{\partial S_1(\beta)}{\partial\beta}\right)^2-48e^{  -2\sqrt{3}\Omega}=0,
\label{sks}
\end{equation}
solving Eq.(\ref{sks}) one gets the functions $S_1,S_2$, and calculate the temporal evolution. First we fix the value of $N(t)=24e^{-\sqrt{3}\beta-2\sqrt{3}\Omega}$, by using (\ref{mks}) and  the definition for the momenta
$
\Pi_\beta=\frac{d S_1(\beta)}{d\beta}$ and $\Pi_\Omega=\frac{d S_2(\Omega)}{d\Omega}$ we obtain the classical solutions
\begin{eqnarray}
\Omega(t)&=&\frac{1}{2\sqrt{3}}\ln\left[\frac{48}{P^2_{\beta_0}}\cosh^2\left(2\sqrt{3}P_{\beta_0}(t-t_0)\right)\right],\nonumber\\
\beta(t)&=&\beta_0+2P_{\beta_0}(t-t_0),
\end{eqnarray}
where $\beta_0$ and $P_{\beta_0}$ are the initial conditions. These solutions are the same that we get by solving the field equations of general relativity.
%%%%%%%%%%%%%%%%%%%%%%%%%%%%
\subsection{FRW cosmology with scalar field and $\Lambda$}
The next set of examples correspond to  homogeneous and isotropic
universes, the so called FRW universe coupled to a scalar field and cosmological constant. The FRW metric is given by:
\begin{equation}
ds^2= -N^2dt^2 + e^{2\alpha(t)}\left[\frac{dr^2}{1-{\rm k} r^2}
+r^2 (d\vartheta^2 + sin^2 \vartheta d\varphi^2) \right] \, ,
\end{equation}
where $a(t)=e^{\alpha(t)}$ is the scale factor, $N(t)$ is the lapse function, and
$\rm k$ is the curvature constant that takes the values $\rm 0,+1,-1$, which
correspond to a flat, closed and open universes, respectively. 
The Lagrangian we are to work on is composed by the gravity sector and the matter sector, which for the FRW universe endowed with a scalar field and cosmological constant $\Lambda$ is
\begin{equation}
{\cal L}_{tot}= {\cal L}_g +{\cal L}_{\phi} =e^{3\alpha} \left[6 \frac{ \dot{\alpha}^2}{N} -
\frac{1}{2} \frac{\dot{\phi^2}}{N} - 
N \left(2 \Lambda + 6 k e^{-2\alpha} \right) \right]\, ,
\label{lagra}
\end{equation}
the corresponding canonical momenta are 
\begin{equation}
 \Pi_\alpha= \frac {\partial {\cal L}}{\partial \dot{\alpha} }=
12 e^{3\alpha}\frac{\dot{\alpha}}{N}~ , \qquad  \Pi_\phi= \frac {\partial {\cal L}}{\partial \dot{\phi} }- e^{3\alpha}\frac{\dot{\phi}}{N}.
 \label{pp}
\end{equation}
Proceeding as before the WDW equation is obtained from the classical Hamiltonian. By the variation of (\ref{lagra}) with respect to $N$,
${\partial {\cal L}}/{\partial N}= 0$, implies the well-known result $\rm {\cal H}=0$. 
\begin{equation}
 e^{-3\alpha}N\left[ -\frac{1}{24}\frac{\partial^2}{\partial\alpha^2}+\frac{1}{2}\frac{\partial^2}{\partial\phi^2}+e^{6\alpha}\left(2 \Lambda + 6 {\rm k} e^{-2\alpha} \right) \right]\Psi(\alpha,\phi)=0.
\end{equation}
Now that we have the complete framework and the corresponding WDW equation, we can proceed to study  different cases.\\ 

In table 1 we can see the different cases that we solved \footnote{The case $k\ne0$, $\Lambda\ne0$ does not have a closed analytical solution to the WDW equation.}, all of them are calculated by using the WKB type procedure, these classical solutions can be derived by solving Einstein's field equations. We can expect that this approximation includes all the gravitational degrees of freedom of the particular cosmological model under study. This almost trivial observation is central to the ideas we are presenting in the next section.
\begin{widetext}
\begin{tabular}{|c|c|c|}
\hline
case & Quantum Solution & Classical Solution\\
\hline
$\rm k$=0,~$\Lambda\ne0$ & $\psi=e^{  \pm i\nu\frac{\sqrt{3}}{2}\phi}  K_{i\nu}\left(  4\sqrt{\frac{\Lambda}{3}}e^{3\alpha}  \right) $ & $\phi(t)=\phi_0-P_{\phi_0}t,$\\
&  and $J_{\nu}$ for $\Lambda<0$& $\alpha(t)=\frac{1}{6}\ln{ \left( \frac{P_{\phi_0}^2}{4\Lambda}\right)}
+ \frac{1}{3}\ln{\left(\textrm{sech}\left[\frac{\sqrt{3}}{2}P_{\phi_0}(t-t_0)\right]\right)},$\\
%& &+$ \frac{1}{3}\ln{\left(\textrm{sech}\left[\frac{\sqrt{3}}{2}P_{\phi_0}(t-t_0)\right]\right)},$\\
%& & \\
\hline 
$\rm k\ne$0,~$\Lambda=0$ & $\psi^{(1)}=e^{  \pm i\frac{\nu}{\sqrt{3}}\phi}  K_{i\nu}\left( 6 e^{2\alpha}  \right) $ for $\rm k=1,$ & $\phi(t)=\phi_0-P_{\phi_0}(t-t_0),$\\
%& for $\rm k=1,$& \\
& $\psi^{(2)}=e^{  \pm i\frac{\nu}{\sqrt{3}}\phi}  J_{\nu}\left(  6 e^{2\alpha}  \right) $ for $\rm k=-1$ & $\alpha(t)=\frac{1}{4}\ln{\left[\frac{P^2_{\phi_0}}{12\rm k}\right]} +\frac{1}{2}\ln{\left(\textrm{sech}\left[\frac{1}{\sqrt{3}}P_{\phi_0}(t-t_0)\right]\right)}$\\
%& for $\rm k=-1$& $ +\frac{1}{2}\ln{\left(\textrm{sech}\left[\frac{1}{\sqrt{3}}P_{\phi_0}(t-t_0)\right]\right)}$,\\
%& & \\
\hline
$\rm k\ne$0,~$\Lambda\ne0$ & Unknown & $\phi(t)=\phi_0-P_{\phi_0}(t-t_0),$ $\int\frac{d\alpha(t)}{\sqrt{P_{\phi_0}-2e^{6\alpha}\left ( 2\Lambda+6 \rm k e^{-2\alpha}\right ) }}=\frac{1}{\sqrt{12}}(t-t_0),$\\
& & \\
%& & $\int\frac{d\alpha(t)}{\sqrt{P_{\phi_0}-2e^{6\alpha}\left ( 2\Lambda+6 \rm k e^{-2\alpha}\right ) }}=\frac{1}{\sqrt{12}}(t-t_0),$\\
%& & \\
%& & \\
\hline
\end{tabular}

Table 1: Classical and quantum solutions for the FRW universe coupled to a scalar field $\phi$. For the case $\Lambda\ne0~\rm k\ne0$, the classical solution for the scale factor is given in an implicit expression. We have fixed the lapse function to $N(t)=e^{3\alpha}$.
\end{widetext}
%%%%%%%%%%%%%%%%%%%%%%%%%%%%
%%%%%%% Stringy cosmology
%%%%%%%%%%%%%%%%%%%%%%%%%
\subsection{Stringy quantum cosmology}
Our final example is related to the graceful exit of pre-big bang cosmology \cite{nuevo}, this model is based on the gravi-dilaton effective action in 1+3 dimensions
\begin{equation}
S=-\frac{\lambda_s}{2}\int d^4x\sqrt{-g}e^{-\phi}(R+\partial_\mu\phi\partial^\nu\phi+V),
\end{equation} 
in this expression $\lambda_s$ is the fundamental string length, $\phi$ is the dilaton field with $V$ the possible dilaton potential. Working with an isotropic background, and setting $a(t)=e^{\beta(t)/\sqrt{3}}$, after integrating by parts, we get
\begin{equation}
S=-\frac{\lambda_s}{2}\int d\tau \left (\bar{\phi'}^2-\beta'^2+Ve^{-2\bar{\phi}}\right ),
\end{equation}
we have used the time parametrization \footnote{The prime denotes differentiation respect to $\tau$.} $dt=e^{-\bar{\phi}}d\tau$, the gauge $g_{00}=1$, and defined $\bar{\phi}=\phi-\ln{\int\left ( \frac{d^3x}{\lambda_s^3}\right )-\sqrt{3}\beta}$. From this action we calculate the canonical momenta, $\Pi_\beta=\lambda_s \beta'$ and $\Pi_{\bar{\phi}}=-\lambda_s \bar{\phi'}$. From the classical hamiltonian we find the WDW equation
\begin{equation}
\left[ \frac{\partial^2}{\partial\bar{\phi}^2}-\frac{\partial^2}{\partial\beta^2}+\lambda_s^2 V(\bar{\phi},\beta)e^{-2\bar{\phi}} \right]\Psi(\bar{\phi},\beta)=0,
\end{equation}
in particular for a potential of the form $V(\bar{\phi})=-V_0e^{m\bar{\phi}}$, the quantum solution is
\begin{equation}
\Psi(\bar{\phi},\beta)=e^{  \pm -i\frac{m-2}{2}\nu\beta}  K_{i\nu}\left[ \frac{2\lambda_s\sqrt{V_0}}{m-2}e^{\left ( \frac{m-2}{2}\right )\bar{\phi}}  \right].
\end{equation}
The classical solutions for the scale factor and the dilaton are
\begin{eqnarray}
\bar{\phi}(\tau)&=&\frac{1}{m-2}\ln\left [ \frac{P^2_{\beta_0}}{V_0\lambda^2_s} \textrm{sech}^2\left(\frac{P_{\beta_0}}{2\lambda_s}(m-2)(\tau-\tau_0) \right )\right ], \nonumber\\
\beta(\tau)&=&\beta_0+\frac{P_\beta}{\lambda_s}(\tau-\tau_0),
\end{eqnarray}
for $m=0$ and $m=4$, the solutions have been obtained in \cite{nuevo}, and are used in connection to the graceful exit from pre-big bang cosmology in quantum string cosmology.

%%%%%%%%%%%%%%%%%%%%%%%%%%%
%%%%% WKB NONCOM
%%%%%%%%%%%%%%%%%%%%%%%%%%%
\section{Noncommutative Quantum Cosmology and the WKB type approximation}

In this section we construct noncommutative quantum cosmology for the 
examples presented in the previous section and calculate the classical evolution via a 
WKB type approximation. To get the classical  cosmological solutions would be a very difficult
 task in any model of noncommutative gravity \cite{ncsdg}, as a consequence of the highly nonlinear character of the field equations.
  We will follow the original proposal of noncommutative quantum cosmology that was developed in\cite{ncqc}. 
  This will allow us to get the desired classical solutions.
   The first noncommutative example that we present is the noncommutative KS 
   followed by the noncommutative FRW universe coupled to a scalar field, and finally stringy noncommutative quantum cosmology. 
   First we present in quite a general form, the construction 
   of noncommutative quantum cosmology and the WKB type method to calculate the classical evolution.

 Let us start with a generic form for the commutative WDW equation, 
 this is  defined in the minisuperspace variables $x,y$. As mentioned in \cite{ncqc} a noncommutative deformation 
 of the minisuperspace variables is assumed
\begin{equation}
[x,y]=i\theta,
\label{ncms}
\end{equation}
this noncommutativity \footnote{This commutation relation implies and uncertainty principle giving and absolute minimal distance in minisuperspace, in \cite{alhu} the same conclusion arises by a different approach.} can be formulated in terms of noncommutative minisuperspace functions with the Moyal product of functions
\begin{equation}
f(x,y)\star g(x,y)=f(x,y)e^{i\frac{\theta}{2}\left(\overleftarrow{\partial_x}\overrightarrow{\partial_y}-\overleftarrow{\partial_y}\overrightarrow{\partial_x} \right) }g(x,y).
\end{equation}
Then the noncommutative WDW equation can be written as
\begin{equation}
\rm \left(-\Pi^2_x+\Pi^2_y- V(x,y)\right)\star\Psi(x,y)=0.
\label{modifiednc}
\end{equation}
We know from noncommutative quantum mechanics \cite{gamboa}, that the symplectic structure is modified changing the commutator algebra. It is possible to return to the original commutative variables and usual commutation relations if we introduce the following change of variables
\begin{equation}
x\to x+\frac{\theta}{2}\Pi_y \qquad \textrm{and} \qquad y\to y-\frac{\theta}{2}\Pi_x.
\label{def}
\end{equation}
The efects of the Moyal star product are reflected in the WDW equation, only through the potential 
\begin{equation}
V(x,y)\star\Psi(x,y)=V(x+\frac{\theta}{2}\Pi_y,y-\frac{\theta}{2}\Pi_x),
\label{ncpotential}
\end{equation}
taking this into account and using the usual substitutions $\rm \Pi_{q^\mu}$=$\rm -i \partial_{q^\mu}$ we arrive to
\begin{equation}
\rm  \left[\frac{\partial^2 }{\partial x^2}
 - \frac{\partial^2}{\partial y^2} -V\left(x-i\frac{\theta}{2}\frac{\partial}{\partial y},y+i\frac{\theta}{2}\frac{\partial}{\partial x}\right)\right]\Psi(x,y)=0,
\end{equation}
this is the noncommutative WDW equation (NCWDW) and it's solutions give the quantum description of the noncommutative Universe. We can use the NCWDW to find the temporal evolution of our noncommutative cosmology by a WKB type procedure. For this we propose that the noncommutative wave function has the form $
\Psi_{NC}(\beta,\Omega)\approx e^{i(S_{NC1}(\beta)+S_{NC2}(\Omega))}$,which in the limit
\begin{eqnarray}
&&\left|\frac{\partial^2 S_{NC1}(\beta)}{\partial\beta^2}\right|<<\left (\frac{\partial S_{NC1}(\beta)}{\partial\beta}\right)^2~,\nonumber\\
&&\left|\frac{\partial^2 S_{NC2}(\Omega)}{\partial\Omega^2}\right|<<\left (\frac{\partial S_{NC2}(\Omega)}{\partial\Omega}\right)^2,
\label{ncwkb2}
\end{eqnarray}
yielding the noncommutative Einstein-Hamilton-Jacobi equation (NCEHJ), that gives the solutions to $S_{NC1}$ and $S_{NC2}$. After the identification $\Pi_{x_{NC}}=\frac{\partial(S_{NC1})}{\partial x}$ and $\Pi_{y_{NC}}=\frac{\partial(S_{NC2})}{\partial y}$ together with the definitions of the canonical momenta and Eq.(\ref{def}) we can find the time dependent solutions for $\rm x$ and $\rm y$.

 In the rest of this section we will apply this ideas to the examples that have already been presented.
\subsection{Noncommutative Kantowski-Sachs Cosmology}
Using the method outlined in the preceding paragraphs, applied to Eq.(\ref{kswdw}) we find the NCWDW equation
\begin{equation}
\left[  \frac{\partial^{2}}{\partial\Omega^{2}}-\frac{\partial^{2}}%
{\partial\beta^{2}}-48e^{  -2\sqrt{3}\left(\Omega-i \frac{\theta}{2}\frac{\partial}{\partial\beta}\right)} \right]
\Psi(\Omega,\beta)=0,
\label{ksncwdw}
\end{equation}
assuming that we can write $\Psi(\Omega,\beta)=e^{\sqrt{3}\nu\beta}X(\Omega)$ the equation for $X(\Omega)$ is
\begin{equation}
\left[-\frac{d^2}{d\Omega^2}+48e^{-3i\nu\theta}e^{-2\sqrt{3}\Omega}+3\nu^2\right]X(\Omega)=0,
\end{equation}
then the solution of the NCWDW equation is
\begin{equation}
\Psi(\Omega,\beta)=e^{\pm i\sqrt{3}\nu\beta}K_{i\nu}\left(4e^{-\sqrt{3}\Omega\pm\frac{3}{2}\nu\theta}\right).
\end{equation}
Usually the next step is to construct a ``Gaussian" wave packet and do the physics with the new wave function. This is not necessary for our purposes, as we are interested on the classical solutions, by applying the WKB type method outlined in the previous section.
Using equations (\ref{wkb1}),
and (\ref{wkb2}) we find the solutions for $S_1(\beta)$ and $S_2(\Omega)$ which have the form
\begin{eqnarray}
&&S_1(\beta)=P_{\beta_0}\beta,\\
&&S_2(\Omega)=-\frac{1}{\sqrt{3}}\sqrt{P^2_{\beta_0}-48e^{-\sqrt{3}\theta P_{\beta_0}}e^{-2\sqrt{3}\Omega}}\nonumber\\
&&+\frac{P_{\beta_0}}{\sqrt{3}}{\rm{arctanh}}\left[ \frac{\sqrt{P^2_{\beta_0}-48e^{-\sqrt{3}\theta P_{\beta_0}}e^{-2\sqrt{3}\Omega}}}{P_{\beta_0}}\right],\nonumber
\end{eqnarray}
then the deformation of the momenta provide us with the noncommutative classical solutions
\begin{eqnarray}
\Omega(t)&=&\frac{1}{2\sqrt{3}}\ln\left[\frac{48}{P^2_{\beta_0}}\cosh^22\sqrt{3}P_{\beta_0}(t-t_0)\right]-\frac{\theta}{2}P_{\beta_0},\nonumber\\
\beta(t)&=&\beta_0+2P_{\beta_0}(t-t_0)\nonumber\\
&-&\frac{\theta}{2}P_{\beta_0}{\rm tanh}^2\left[ 2\sqrt{3}P_{\beta_0}(t-t_0)\right],
\end{eqnarray} 
these solutions have already been obtained in\cite{barbosa1}. In that paper the authors deform the symplectic structure at a classical level changing the Poisson brackets.
%%%%%%%%%%%%%%%%%%%%%%%%
\subsection{Noncommutative FRW cosmology with scalar field and $\Lambda$}
We can use the NCWKB type method to FRW universe coupled to a scalar field.
Proceeding as before the corresponding NCWDW equation is  
\begin{equation}
\left[ -\frac{1}{24}\frac{\partial^2}{\partial\alpha^2}+\frac{1}{2}\frac{\partial^2}{\partial\phi^2}+e^{6(\alpha-i\frac{\theta}{2}\frac{\partial}{\partial \phi})}\left(2 \Lambda + 6 \rm k e^{-2(\alpha-i\frac{\theta}{2}\frac{\partial}{\partial \phi})} \right) \right]\Psi=0.
\end{equation}
From the NCWDW equation, we use the method developed in the previous sections and calculate the classical evolution by appliying the NCWKB type method. These results are presented in the next table
\begin{widetext}
\begin{tabular}{|c|c|c|}
\hline
case & NC Quantum Solution & NC Classical Solution\\
\hline
$\rm k$=0,~$\Lambda\ne0$ & $\psi=e^{  \pm i\nu\frac{\sqrt{3}}{2}\phi}  K_{i\nu}\left [ 4\sqrt{\frac{\Lambda}{3}}e^{3\left(\alpha-\frac{3}{2}\nu\theta\right)}  \right] $ & $\phi(t)=\phi_0-P_{\phi_0}t-\sqrt{3}\theta P_{\phi_0}\tanh{\left(\frac{\sqrt{3}}{2}P_{\phi_0}(t-t_0)\right)},$\\
%& & $-\sqrt{3}\theta P_{\phi_0}\tanh{\left(\frac{\sqrt{3}}{2}P_{\phi_0}(t-t_0)\right)},$\\
&  and $J_{\nu}$ for $\Lambda<0$& $\alpha(t)=\frac{\theta}{2}P_{\phi_0}+\frac{1}{6}\ln{ \left( \frac{P_{\phi_0}^2}{4\Lambda}\right)}+\frac{1}{3}\ln{\left(\textrm{sech}\left[\frac{\sqrt{3}}{2}P_{\phi_0}(t-t_0)\right]\right)}$,\\
%& &+$ \frac{1}{3}\ln{\left(\textrm{sech}\left[\frac{\sqrt{3}}{2}P_{\phi_0}(t-t_0)\right]\right)},$\\
%& & \\
\hline 
$\rm k\ne$0,~$\Lambda=0$ & $\psi^{(1)}=e^{  \pm i\frac{\nu}{\sqrt{3}}\phi}  K_{i\nu}\left[ 6 e^{2\left(\alpha-\frac{\theta}{2}\nu\right)}  \right] $ for $\rm k=1$ & $\phi(t)=\phi_0-P_{\phi_0}(t-t_0)-\sqrt{3}\theta P_{\phi_0}\tanh{\left(\frac{P_{\phi_0}}{\sqrt{3}}(t-t_0)\right)},$\\
%& for $\rm k=1$& $-\sqrt{3}\theta P_{\phi_0}\tanh{\left(\frac{P_{\phi_0}}{\sqrt{3}}(t-t_0)\right)},$\\
& $\psi^{(2)}=e^{  \pm i\nu/\sqrt{3}\phi}  J_{\nu}\left[  6 e^{2\left(\alpha-\frac{\theta}{2}\nu\right)}  \right] $, for $\rm k=-1$ & $\alpha(t)=\frac{\theta}{2}P_{\phi_0}+\frac{1}{4}\ln{\left[\frac{P^2_{\phi_0}}{12\rm k}\right]}+\frac{1}{2}\ln{\left(\textrm{sech}\left[\frac{1}{\sqrt{3}}P_{\phi_0}(t-t_0)\right]\right)}$\\
%& for $\rm k=-1$& $ +\frac{1}{2}\ln{\left(\textrm{sech}\left[\frac{1}{\sqrt{3}}P_{\phi_0}(t-t_0)\right]\right)}$,\\
\hline
$\rm k\ne$0,~$\Lambda\ne0$ & Unknown & $\phi(t)=\phi_0-P_{\phi_0}(t-t_0)+ 6\theta\int e^{6\alpha}\left(\Lambda+2e^{-2\alpha}\right)dt,$\\
%& & + $6\theta\int e^{6\alpha}\left(\Lambda+2e^{-2\alpha}\right)dt,$\\
& & $\int\frac{d\alpha(t)}{\sqrt{P_{\phi_0}-2e^{6\alpha+3\theta P_{\phi_0}}\left ( 2\Lambda+6\rm k e^{-2\alpha-\theta P_{\phi_0}}\right ) }}=\frac{1}{\sqrt{12}}(t-t_0),$\\
%& & \\
\hline
\end{tabular}

Table 2: Classical and quantum solutions for noncommutative FRW universe coupled to a scalar field. For these models noncommutativity is introduced in the gravitational and matter sectors. As in the commutative scenario, for $\Lambda\ne0$ and $\rm k\ne0$ the noncommutative classical solution is given in an implicit form, and there is not a closed analytical quantum solutions. As in the commutative case we have fixed the value of the lapse function $N(t)=e^{3\alpha}$.
\end{widetext}
%%%%%%%%%%%%%%%%%%%%%%%%%%%%%%%%%
\subsection{Stringy noncommutative quantum cosmology}
As in the previous examples we introduce the noncommutative relation $[\bar{\phi},\beta]=i\theta$, and from the classical hamiltonian we find the NCWDW equation
\begin{equation}
\left[ \frac{\partial^2}{\partial\bar{\phi}^2}-\frac{\partial^2}{\partial\beta^2}-\lambda_s^2 V(\bar{\phi},\beta)e^{(m-2)(\bar{\phi}-i\frac{\theta}{2}\frac{\partial}{\partial\beta})} \right]\Psi(\bar{\phi},\beta)=0.
\end{equation}
The noncommutative wave function is
\begin{equation}
\Psi(\bar{\phi},\beta)=e^{  \pm -i\frac{m-2}{2}\nu\beta}  K_{i\nu}\left[ \frac{2\lambda_s\sqrt{V_0}}{m-2}e^{(m-2)\left (\bar{\phi}\mp \frac{m-2}{4}\theta\nu \right )}  \right],
\end{equation}
using the NCWKB type method the classical solutions for the noncommutative stringy cosmology are
\begin{eqnarray}
\bar{\phi}(\tau)&=&\frac{1}{m-2}\ln\left [ \frac{P^2_{\beta_0}}{V_0\lambda^2_s } \textrm{sech}^2\left(\frac{P_{\beta_0}}{2\lambda_s}(m-2)(\tau-\tau_0) \right )\right ]\nonumber\\
&-&\frac{\theta}{2}P_{\beta_0}, \nonumber\\
\beta(\tau)&=&\beta_0+\frac{P_\beta}{\lambda_s}(\tau-\tau_0)\nonumber\\
&+&\theta\frac{P_{\beta_0}}{2}{\textrm{tanh}}\left[ \frac{P_{\beta_0}}{2\lambda_s}(m-2)(\tau-\tau_0)\right],
\end{eqnarray}
the classical evolution for string cosmology can be calculated for $m=0$ and $m=4$. 
An interesting issue concerns the $B$ field that is turned off in the string cosmology model \cite{nuevo}  
and does  not contribute to the effective action. In open string theory, however noncommutativity arises 
precisely in the low energy limit of string theory in the presence of a constant $B$ field. 
The $\theta$ parameter  we have introduced in the minisuperspace could then be understood as a
 kind of B-field related with the Neveu-Schwarz B-field.
 
\section{Conclusions and Outlook}
In this paper we have presented the NCWKB type method for noncommutative quantum cosmology and with this procedure, found the noncommutative classical solutions for several noncommutative quantum cosmological models.

Noncommutativity is a proposal that originally emerged at the quantum level,  moreover space-time noncommutativity has as a consequence that the fields do not commute \cite{ref3,Jurco:2001rq,ncsdg}, by this reason we incorporate noncommutativity
in the minisuperspace variables in a similar manner as it is considered in standard quantum mechanics. By means of the WKB approximation on the corresponding NCWDW equation, one gets the noncommutative generalized Einstein-Hamilton-Jacobi equation (NCEHJ), from which the classical evolution of the noncommutative model is obtained. The examples we studied were the Kantowski-Sachs cosmological model, the FRW universe with cosmological constant and coupled to a scalar field, and a string quantum cosmological model.
In the commutative scenario, the classical solutions found from the WKB-type method are solutions to the corresponding Einsteins field equations. 
Due to the complexity of the noncommutative theories of gravity \cite{ncsdg}, classical solutions to the noncommutative field equations are almost impossible to find, but in the approach of noncommutative quantum cosmology and by means of the WKB-type procedure, they can be easily constructed.  Also the quantum evolution of the system is not needed to find the classical behavior, from table 2 we can see that for the case $\Lambda\ne 0$ and $\rm k\ne 0$ the wave function can not be analitacally calculated, but still the noncommutative effects can be incorporated and the classical evolution is found implicitly. 
This procedure gives a straightforward algorithm to incorporate noncommutative effects to cosmological models.
 In this approach the effects of noncommutativity are encoded in the potential through the Moyal product of functions 
  Eq. (\ref{ncpotential}). We only need the NCWDW equation and the approximations (\ref{wkb2}),
   to get the NCEHJ and from it, the noncommutative classical behavior can easily be constructed.
As already mentioned, in \cite{brand} the effects of noncommutativity were studied in connection with inflation, 
but the noncommutative deformation was only done in the matter sector neglecting the gravity sector.
 The procedure developed here has the advantage that we can implement noncommutativity in both sectors 
 in a straightforward way and find the classical solutions (i.e. inflationary models). 
 These ideas are being explored and will be reported elsewhere.

%%%%

\section*{Acknowledgments}
We will like to thank M. P. Ryan for  enlightening discussions on quantum cosmology and G. Garc\'{\i}a for comments about the manuscript. This work was partially supported by CONACYT grants  47641 and 51306,
 and PROMEP grants UGTO-CA-3 and PROMEP-PTC-085.


\begin{thebibliography}{99}
\bibitem{snyder} H. Snyder, Phys. Rev. {\bf 71}, 38 (1947).
\bibitem{connes}  A. Connes, M. R. Douglas, and A. Schwarz, {JHEP} {\bf %
9802:003} (1998).
\bibitem{ref3}  N. Seiberg and E. Witten, {JHEP} {\bf 9909:032} (1999).
\bibitem{Jurco:2001rq}
  B.~Jurco, L.~Moller, S.~Schraml, P.~Schupp and J.~Wess,
  %``Construction of non-Abelian gauge theories on noncommutative spaces,''
  Eur.\ Phys.\ J.\ C {\bf 21}, 383 (2001),hep-th/0104153.
  %%CITATION = HEP-TH 0104153;%%
\bibitem{ncsdg}
  H.~Garc\'ia-Compe\'an, O.~Obreg\'on, C.~Ram\'irez and M.~Sabido,
  %``Noncommutative self-dual gravity,''
  Phys.\ Rev.\ D {\bf 68} (2003) 044015;  H.~Garc\'ia-Compe\'an, O.~Obreg\'on, C.~Ram\'irez and M.~Sabido,
  %``Noncommutative topological theories of gravity,''
  Phys.\ Rev.\ D {\bf 68} (2003) 045010; M. Ba\~{n}ados, O. Chandia, N. Grandi, F.A. Schaposnik
and G.A. Silva, Phys. Rev. D {\bf 64} (2001) 084012; H. Nishino and S. Rajpoot, Phys. Lett. B {\bf 532} (2002)
334; V.P. Nair, ``Gravitational Fields on a Noncommutative
Space", hep-th/0112114; S. Cacciatori, D. Klemm, L. Martucci and D. Zanon, Phys. Lett. B {\bf 536} (2002) 101; S. Cacciatori, A.H. Chamseddine, D. Klemm, L. Martucci,
W.A. Sabra and D. Zanon, ``Noncommutative Gravity in Two Dimensions", hep-th/0203038; Y. Abe and V.P. Nair,``Noncommutative Gravity: Fuzzy Sphere and Others", hep-th/0212270; M.A. Cardella and Daniela Zanon, ``Noncommutative Deformation of
Four-dimensional Einstein Gravity", hep-th/0212071; A.H. Chamseddine, ``Invariant Actions for Noncommutative
Gravity'', hep-th/0202137; J.W. Moffat, {Phys. Lett.} B {\bf 491} (2000) 345; {Phys.
Lett.} B {\bf 493} (2000) 142; V.~O.~Rivelles,
  %``Noncommutative field theories and gravity,''
  Phys.\ Lett.\ B {\bf 558} (2003) 19; E.~Harikumar and V.~O.~Rivelles,
  %``Noncommutative gravity,''
  arXiv:hep-th/0607115.

\bibitem{min}
  S.~Minwalla, M.~Van Raamsdonk and N.~Seiberg,
  %``Noncommutative perturbative dynamics,''
  JHEP {\bf 0002}, 020 (2000).
  %%CITATION = HEP-TH 9912072;%%
\bibitem{doug}
  M.~R.~Douglas and N.~A.~Nekrasov,
  %``Noncommutative field theory,''
  Rev.\ Mod.\ Phys.\  {\bf 73}, 977 (2001).
  %%CITATION = HEP-TH 0106048;%%

\bibitem{ryan}M.P. Ryan, in: {\it Hamiltonian Cosmology} (Springer, Berlin, 1972).
\bibitem{maccallum} M. MacCallum, in: {\it General Relativity: An Einstein Centenary 
Survey},   edited by S. Hawking and W. Israel
  (Cambridge University Press, Cambridge, England, 1979).
\bibitem{romero}
 J.~M.~Romero and J.~A.~Santiago,
  %``Cosmological constant and noncommutativity: A Newtonian approach,''
  Mod.\ Phys.\ Lett.\ A {\bf 20} (2005) 78.
  %%CITATION = HEP-TH 0310266;%%
 \bibitem{brand}
   R.~Brandenberger and P.~M.~Ho,
  Phys.\ Rev.\ D {\bf 66} (2002) 023517; Q.~G.~Huang and M.~Li,
  %``Power spectra in spacetime noncommutative inflation,''
  Nucl.\ Phys.\ B {\bf 713} (2005) 219; Q.~G.~Huang and M.~Li,
  %``CMB power spectrum from noncommutative spacetime,''
  JHEP {\bf 0306} (2003) 014; H.~s.~Kim, G.~S.~Lee, H.~W.~Lee and Y.~S.~Myung,
  %``Second-order corrections to noncommutative spacetime inflation,''
  Phys.\ Rev.\ D {\bf 70} (2004) 043521; H.~s.~Kim, G.~S.~Lee and Y.~S.~Myung,
  %``Noncommutative spacetime effect on the slow-roll period of inflation,''
  Mod.\ Phys.\ Lett.\ A {\bf 20} (2005) 271; D.~J.~Liu and X.~Z.~Li,
  %``Cosmological perturbations and noncommutative tachyon inflation,''
  Phys.\ Rev.\ D {\bf 70} (2004) 123504.
  \bibitem{ncqc}
H.~Garc\'ia-Compe\'an, O.~Obreg\'on and C.~Ram\'irez,
   %``Noncommutative quantum cosmology,''
  Phys.\ Rev.\ Lett.\  {\bf 88}, 161301 (2002).

    \bibitem{gamboa}
  J.~Gamboa, M. Loewe and J. C. Rojas., Phys.\ Rev.\ D {\bf 64},067901;M. Chaichian, M. M. Sheikh-Jabbari, and  A. Tureanu., Phys. \ Rev. Lett. {\bf 86}, 2716.

\bibitem{nuevo}
M.~Gasperini, J.~Maharana and G.~Veneziano,
  %``Graceful exit in quantum string cosmology,''
  Nucl.\ Phys.\ B {\bf 472} (1996) 349. 
 \bibitem{alhu}
 D.~V.~Ahluwalia,
   %``Quantum violation of the equivalence principle and gravitationally
  %modified de Broglie relation,''
  Phys.\ Lett.\ A {\bf 275}, 31 (2000).
\bibitem{barbosa1}
  G.~D.~Barbosa and N.~Pinto-Neto,
  %``Noncommutative geometry and cosmology,''
  Phys.\ Rev.\ D {\bf 70} (2004) 103512.
  %%CITATION = HEP-TH 0407111;%%
  %\cite{Pimentel:2004jv}
\bibitem{pimentel}  
L.~O.~Pimentel and C.~Mora,
  %``Noncommutative quantum cosmology,''
  Gen.\ Rel.\ Grav.\  {\bf 37}, 817 (2005).

\end{thebibliography}
\end{document}